\documentclass[preprintnumbers,amsmath,amssymb,floatfix,10pt,prd,onecolumn,
superscriptaddress,nofootinbib]{revtex4}
\usepackage{latexsym}
\usepackage{epsfig}
\usepackage{epstopdf}
\usepackage{graphicx}
\usepackage{amssymb}
\usepackage{amsmath}
\usepackage{dcolumn}
\usepackage{bm}
\usepackage{color}
\usepackage{comment}
\begin{document}

\title{\bf Existence of stable wormholes on a noncommutative-geometric background in modified gravity}

\author{M. Zubair}
\email{mzubairkk@gmail.com;drmzubair@ciitlahore.edu.pk}\affiliation{Department of Mathematics, COMSATS, Institute of Information
Technology Lahore, Pakistan}

\author{G. Mustafa}
\email{gmustafa3828@gmail.com}\affiliation{Department of Mathematics, COMSATS, Institute of Information Technology Lahore,
Pakistan}

\author{Saira Waheed}
\email{swaheed@pmu.edu.sa}\affiliation{Prince Mohammad Bin Fahd University, Al Khobar, 31952 Kingdom of Saudi Arabia.}

\author{G. Abbas}
\email{abbasg91@yahoo.com}\affiliation{Department of Mathematics,
The Islamia University of Bahawalpur, Bahawalpur, Pakistan.}

\begin{abstract}

In this paper, we discuss spherically symmetric wormhole solutions
in $f(R,T)$ modified theory of gravity by introducing well-known
non-commutative geometry in terms of Gaussian and Lorentizian
distributions of string theory. For some analytic discussion, we
consider an interesting model of $f(R,T)$ gravity defined by
$f(R,T)=f_{1}(R)+\lambda T$. By taking two different choices for the
function $f_{1}(R)$, that is, $f_{1}(R)=R$ and $f_{1}(R)=R+\alpha
R^{2}+\gamma R^{n}$, we discuss the possible existence of wormhole
solutions. In the presence of non-commutative Gaussian and
Lorentizian distributions, we get exact and numerical solutions for
both these models. By taking appropriate values of the free
parameters, we discuss different properties of these wormhole models
analytically and graphically. Further, using equilibrium condition,
it is found that these solutions are stable. Also, we discuss the
phenomenon of gravitational lensing for the exact wormhole model
and it is found that the deflection angle diverges at wormhole throat.\\

\end{abstract}

\maketitle

\date{\today}

\section{Introduction}
In modern cosmology, the phenomenon of accelerated cosmic expansion
and its possible causes, as confirmed by numerous astronomical
probes, have become a center of interest for the researchers
\cite{1}-\cite{3}. In this respect, the first attempt was made by
Einstien, by introducing a well-known $\Lambda CDM$ model but in
spite of its all beauty and success, this model cannot be proved as
problem free \cite{4}. Later on, a bulk of different proposals have
been presented by the researchers that can be grouped into two
kinds: modified matter proposals and modified curvature proposals.
Tachyon model, quintessence, Chaplygin gas and its different
versions, phantom, quintom etc. are all obtained by introducing some
extra terms in matter section and hence are members of the modified
matter proposal group \cite{5}. The other idea is to modify the
curvature sector of Einstein's general relativity (GR) by including
some extra degrees of freedom there. One of the primary alterations
was the speculation of the Einstein-Hilbert lagrangian density with
an arbitrary function $f(R)$ instead of Ricci scalar $R$. This
theory has been widely used in literature \cite{6} to examine the
dark energy (DE) and its resulting speedy cosmic expansion.
Moreover, $f(R)$ theory of gravitation provides a unified picture of
early stages of cosmos (inflation) as well as the late stages of
accelerated cosmos. Some other well-known examples include
Brans-Dicke gravity, generalized scalar-tensor theory, $f(\tau)$
gravity, where $\tau$ is a torsion, Gauss-Bonnet gravity and its
generalized forms like $f(G)$ gravity, $f(R,G)$ gravity, and
$f(\tau, \tau_G)$ theory, etc. \cite{7}.

Another significant modification of Einstein gravity namely $f(R,T)$
gravity was proposed by Harko et al. \cite{8} almost five years ago.
In this formulation, a generic function $f(R,T)$, representing the
coupling of Ricci scalar and energy-momentum tensor trace, replaces
the Ricci scalar $R$ for the possible modification of curvature
sector. Using metric formalism, they derived the associated field
equations for some specific cases. In \cite{9}, some interesting
cosmological $f(R,T)$ models have been developed by employing
various scenarios namely, auxiliary scalar field, dark energy models
and anisotropic universe models.  In literature \cite{12}, different
cosmological applications of $f(R,T)$ gravity have been discussed
like energy conditions, thermodynamics, exact and numerical
solutions of field equations with different matter content, phase
space perturbation, compact stars and stability of collapsing
objects etc.

The existence and construction of wormhole solutions is one of the
most fascinating topics in modern cosmology. Wormholes are
topological passage like structures connecting two distant parts of
the same universe or different universes together through a shortcut
called tunnel or bridge. Generally, in nature, wormholes are
categorized into two sorts namely static wormholes and dynamic
wormholes \cite{13}. For the development of wormhole structures, an
exotic fluid (hypothetical form of matter) is required which
violates the null energy condition (NEC) in GR. This violation of
energy condition is regarded as one of the basic requirements for
wormhole construction. The existence of wormhole solutions in GR has
always been a great challenge for the researchers. Although GR
allows the existence of wormholes but it is necessary to first
modify the matter sector by including some extra terms (as the
ordinary matter satisfies the energy bounds and hence violates the
basic criteria for wormhole existence). These extra terms are
responsible for energy bound violation and hence permits the
existence of wormhole in GR. In 1935, Einstein and Rosen \cite{14}
discussed the mathematical criteria of wormholes in GR and they
obtained the wormhole solutions known as Lorentzian wormholes or
Schwarzchild wormholes. In 1988, it was shown \cite{15} that
wormholes could be large enough for humanoid travelers and even
permit time travel. In literature \cite{16,17}, numerous authors
constructed wormholes by including different types of exotic matter
like quintom, scalar field models, non-commutative geometry and
electromagnetic field etc. and obtained different interesting and
physically viable results. Some important and interesting results
regarding the stable wormhole solutions without inclusion of any
exotic matter are discussed in \cite{a}. In a recent paper \cite{b},
the existence of wormhole solutions and its different properties in
$f(R,T)$ theory gravity has been discussed.


``On a D-brane, the coordinates may be treated as non-commutative
operators", this is one of the most interesting aspect of
non-commutative geometry of string theory that provides a
mathematical way to explore some important concepts of quantum
gravity \cite{18}. Basically, non-commutative geometry is an effort
to construct a unified platform where one can take the spacetime
gravitational forces as a combined form of weak and strong forces
with gravity. Non-commutativity has an important feature of
replacing point-like structures by smeared objects and hence
corresponds to spacetime discretization which is due to the
commutator defined by $[x^{\alpha}, x^{\beta}] =
i\theta^{\alpha\beta}$, where $\theta^{\alpha\beta}$ is an
anti-symmetric second-order matrix. This smearing effect can be
modeled by including Gaussian distribution and Lorentizian
distribution of minimal length $\sqrt{\theta}$ instead of the Dirac
delta function. The spherically symmetric, static particle like
gravitational source representing Gaussian distribution of
non-commutative geometry with total mass $M$ has energy density
given by \cite{19}
\begin{equation*}\label{n1}
\rho(r)=\frac{M}{(4\pi \theta)^{\frac{3}{2}}
}e^{-\frac{r^{2}}{4\theta}},
\end{equation*}
while with reference to Lorentzian distribution, we can take the
density function of particle-like mass $M$ as follows
\begin{equation*}\label{n2}
\rho(r)=\frac{M \sqrt{\theta}}{\pi^{2}(r^{2}+\theta)^{2}}.
\end{equation*}
Here total mass $M$ can be considered as wormhole, a type of
diffused centralized object and clearly, $\theta$ is the
noncommutative parameter. The Gaussian distribution source has been
utilized by Sushkov to model phantom-energy upheld wormholes
\cite{19a}. Also, Nicolini and Spalluci \cite{20} used this
distribution to demonstrate physical impacts of short-separation
changes of non-commutative coordinates in the investigation of black
holes.

Being motivated from this literature, in this manuscript, we will
construct spherically symmetric static wormholes in the presence of
curvature matter coupling with non-commutative geometry. In the next
section, we will describe the basic mathematical formulation of
$f(R,T)$ gravity and the corresponding field equations for static
spherically symmetric spacetime. In section \textbf{III}, we shall
discuss the wormhole solutions for both Gaussian and Lorentzian
distributions of non-commutative geometry by taking linear model of
$f(R,T)$ gravity, i.e., $f(R,T)=R+\lambda T$. Section \textbf{IV}
provides wormhole solutions for both these distributions of
non-commutative geometry where the model $f(R,T)=R+\alpha
R^{2}+\gamma R^{n}+\lambda T$ will be taken into account. In section
\textbf{V}, the stability of these obtained wormhole solutions will
be discussed through graphs. Section \textbf{VI} will be devoted to
investigate the gravitational lensing phenomenon for the exact model
of section \textbf{III} by exploring deflection angle at the
wormhole throat. Last section will summarize the whole discussion by
highlighting the major achievements.

\section{Field Equations of $f(R,T)$ Gravity and Spherically Symmetric Wormhole Geometry}

In this section, we shall discuss the basic formulation of $f(R,T)$
gravity and its corresponding field equations for spherically
symmetric spacetime in the presence of ordinary matter. For this
purpose, we take the following action of this modified gravity
\cite{8}:
\begin{equation}\label{1}
S=\frac{1}{16\pi}\int f(R,T)\sqrt{-g}d^{4}x+\int L_{m}\sqrt{-g}d^{4}x,
\end{equation}
where $f(R,T)$ is an arbitrary function of Ricci scalar $R$ and the
trace of energy-momentum tensor $T=g^{\mu\nu}T_{\mu\nu}$. Here
$L_{m}$ represents the Lagrangian density of ordinary matter. By
taking variation of the above action, we have the following set of
equations:
\begin{eqnarray}\nonumber
8\pi
T_{\mu\nu}-f_{T}(R,T)T_{\mu\nu}-f_{T}(R,T)\Theta_{\mu\nu}&=&f_{R}(R,T)R_{\mu\nu}
-\frac{1}{2}f(R,T)g_{\mu\nu}\\\label{2}&+&(g_{\mu\nu}\Box-\nabla_{\mu}\nabla_{\nu})f_{R}(R,T).
\end{eqnarray}
By contracting the above equation, we have a relation between Ricci
scalar $R$ and the trace $T$ of the energy momentum tensor as
follows
\begin{eqnarray}\label{3}
8\pi T-f_{T}(R,T)T-f_{T}(R,T)\Theta=f_{R}(R,T)R+3\Box f_{R}(R,T)-2f(R,T).
\end{eqnarray}
These two equations involves covariant derivative and d'Alembert
operator denoted by $\nabla$ and $\Box$, respectively. Furthermore,
$f_{R}(R,T)$ and $f_{T}(R,T)$ correspond to the function derivatives
with respect to $R$ and $T$, respectively. Also, the term
$\Theta_{\mu\nu}$ is defined by
\begin{equation}\nonumber
\Theta_{\mu\nu}=\frac{g^{\alpha\beta}\delta T_{\mu\nu}}{\delta
g^{\mu\nu}}=-2T_{\mu\nu}+g_{\mu\nu}L_{m}-2g^{\alpha\beta}\frac{\partial^{2}L_{m}}{\partial
g^{\mu\nu}\partial g^{\alpha\beta}}.
\end{equation}

The energy-momentum tensor for anisotropic fluid is given by
\begin{equation}\nonumber
T_{\mu\nu}=(\rho+p_{t})V_{\mu}V_{\nu}-p_{t}g_{\mu\nu}+(p_{r}-p_{t})\chi_{\mu}\chi_{\nu},
\end{equation}
where $V_{\mu}$ is the 4-velocity vector of the fluid given by
$V^{\mu}=e^{-a}\delta^{\mu}_{0}$ and
$\chi^{\mu}=e^{-b}\delta^{\mu}_{1}$ which satisfy the relations:
$V^{\mu}V_{\mu}=-\chi^{\mu}\chi_{\mu}=1$. Here we choose
$L_{m}=\rho$, which leads to following expression for
$\Theta_{\mu\nu}$:
\begin{equation}\nonumber
\Theta_{\mu\nu}=-2T_{\mu\nu}-\rho g_{\mu\nu}.
\end{equation}
We relate the trace equation (\ref{3}) with equation (\ref{2}), then
Einstein field equations take the form given by
\begin{eqnarray}\nonumber
f_{R}(R,T)G_{\mu\nu}&=&(8\pi+f_{T}(R,T))T_{\mu\nu}+[\nabla_{\mu}\nabla_{\nu}f_{R}(R,T)\\\label{4}&-&
\frac{1}{4}g_{\mu\nu}\{(8\pi+f_{T}(R,T))T+\Box
f_{R}(R,T)+f_{R}(R,T)R)\}].
\end{eqnarray}

The spherically symmetric wormhole geometry is defined by the
spacetime:
\begin{equation}\label{5}
ds^2=-e^{2\Phi(r)}dt^2+\frac{dr^2}{1-b(r)/r}+r^{2}(d\theta^{2}+sin^{2}\theta d\Phi^{2}),
\end{equation}
where $\Phi(r)$ and $b(r)$ both are functions of radial coordinate $r$ and
represent redshift and shape functions, respectively \cite{15, 21}. In the
subsequent discussion, we shall assume the red shift function to be constant,
i.e., $\Phi'(r)=0$. Here the radial coordinate $r$ is non-monotonic as it
decreases from infinity to a minimum value $r_{0}$, representing the location
of wormhole throat, i.e., $b(r_{0})=r_{0}$, then it increases back from
$r_{0}$ to infinity. The most important condition for wormhole existence is
the flaring out property where the shape function satisfies the inequality:
$(b-b^{'}r)/b^{2}>0$, while at the wormhole throat, it satisfies
$b(r_{0})=r_{0}$. Further, the property $b^{'}(r_{0})<1$, is also a necessary
condition to be satisfied for the wormhole solutions. Basically these
conditions lead to NEC violation in classical GR. Furthermore, another
condition that needs to be satisfied for wormhole solutions is $1-b(r)/r>0$.
These all conditions collectively provide a basic criteria for the existence
of a physically realistic wormhole model.

In order to find the relations for $\rho,~p_{r}$ and $p_{t}$, we
substitute the corresponding quantities for the metric (\ref{5}) in
the equation (\ref{4}) and then by rearranging the resulting
equations, we have
\begin{eqnarray}\label{6}
\frac{b^{'}}{r^2}&=&\frac{(8\pi+f_{T}(R,T))}{f_{R}(R,T)}\rho+\frac{H}{f_{R}(R,T)},\\\nonumber
-\frac{b}{r^3}&=&\frac{(8\pi+f_{T}(R,T))}{f_{R}(R,T)}p_{r}+\frac{1}{f_{R}(R,T)}(1-\frac{b}{r})[(f^{''}_{R}(R,T)\\\label{7}&-&f^{'}_{R}(R,T)
\frac{(b^{'}r-b)}{2r^{2}(1-b/r)})]-\frac{H}{f_{R}(R,T)},\\\nonumber
-\frac{b^{'}r-b}{2r^3}&=&\frac{(8\pi+f_{T}(R,T))}{f_{R}(R,T)}p_{t}+\frac{1}{f_{R}(R,T)}(1-\frac{b}{r})\frac{f^{'}_{R}(R,T)}{r}\\\label{8}
&-&\frac{H}{f_{R}(R,T)},
\end{eqnarray}
where
\begin{equation}\label{9}
H=H(r)=\frac{1}{4}(f_{R}(R,T)R+\Box f_{R}(R,T)+(8\pi+f_{T}(R,T))T).
\end{equation}
The curvature scalar $R$ is given by
\begin{equation}\label{10}
R=\frac{2b^{'}}{r^2}
\end{equation}
and $\Box f_{R}(R,T)$ has the following expression
\begin{equation}\label{11}
\Box f_{R}(R,T)=(1-\frac{b}{r})[f^{''}_{R}(R,T)-f^{'}_{R}(R,T)
\frac{(b^{'}r-b)}{2r^{2}(1-b/r)}+\frac{2f^{'}_{R}(R,T)}{r}].
\end{equation}
Since the above system, involving higher-order derivatives with many
unknowns, is very complicated to solve for the quantities
$\rho,\;p_{r}$ and $p_{t}$ therefore, for the sake of simplicity in
calculations, we assume a particular form of the function $f(R,T)$
given by the relation $f(R,T)=f_{1}(R)+f_{2}(T)$ with
$f_{2}(T)=\lambda T$, where $\lambda$ is a coupling parameter. After
inserting this form of $f(R,T)$ and then by simplifying the
corresponding equations (\ref{6})-(\ref{8}), we get
\begin{eqnarray}
\rho&&=\frac{b^{'}f_{R}}{r^{2}(8\pi+\lambda)},\label{12}\\
p_{r}&&=-\frac{b f_{R}
}{r^3(8\pi+\lambda)}+\frac{f^{'}_{R}}{2r^{2}(8\pi+\lambda)}(b^{'}r-b)-(1-\frac{b}{r})\frac{f^{''}_{R}}{(8\pi+\lambda)},\label{13}\\
p_{t}&&=-\frac{f^{'}_{R}}{r(8\pi+\lambda)}(1-\frac{b}{r})+\frac{f_{R}}{2r^{3}(8\pi+\lambda)}(b^{'}r-b).\label{14}
\end{eqnarray}

\section{Wormhole Solutions: Gaussian and Lorentzian Distributions for $f_1(R)=R$ model}

In this section, we shall consider a specific and interesting $f(R)$
model \cite{22} that is given by the linear function of Ricci
scalar:
\begin{equation}\label{15}
f_1(R)=R.
\end{equation}
Using this relation in Eqs.(\ref{12})-(\ref{14}) and after doing
some simplifications, we get the following set of field equations:
\begin{eqnarray}\label{16}
\rho&&=\frac{b^{'}}{r^{2}(8\pi+\lambda)},\\\label{17}
p_{r}&&=-\frac{b }{r^3(8\pi+\lambda)},\\\label{18}
p_{t}&&=\frac{(b-b^{'}r)}{2r^{3}(8\pi+\lambda)}.
\end{eqnarray}
Here we include the smearing effect mathematically by substituting
Gaussian distribution of insignificant width $\sqrt{\theta}$ in the
place of Dirac-delta function, where $\theta$ is a noncommutative
parameter of Gaussian distribution. Here we consider the mass
density of a static, spherically symmetric, smeared, particle-like
gravitational source given by
\begin{equation}\label{19}
\rho(r)=\frac{M}{(4\pi \theta)^{\frac{3}{2}}
}e^{-\frac{r^{2}}{4\theta}}.
\end{equation}
The particle mass $M$, rather than of being splendidly restricted at
the point, diffused on a region of direct estimate $\sqrt{\theta}$.
This is because of fact that the uncertainty is encoded in the
coordinate commutator.

Comparing equations (\ref{16}) and (\ref{19}), and then by solving
the resulting differential equation, we get the shape function
$b(r)$ in terms of error function as follows
\begin{equation}\label{20}
b(r)=m_{0}[-2r\theta e^{-\frac{r^{2}}{4\theta}}+2
\theta^{\frac{3}{2}}\pi
^{\frac{1}{2}}erf\{\frac{r}{2\sqrt{\theta}}\}+C_1],
\end{equation}
where
\begin{equation*}
m_{0}= \frac{M(8\pi+\lambda)}{8\pi^\frac{3}{2}\theta^\frac{3}{2}}
\end{equation*}
and
\begin{equation*}
erf(\theta)=\frac{2}{\sqrt{\pi}}\int^{\theta}_{0}e^{-t^{2}}dt.
\end{equation*}
Here $C_1$ is a constant of integration. Also, $\lambda\neq-8\pi$
which clearly leads to $b(r)=0$. Using equation (\ref{20}) in
(\ref{16})-(\ref{18}), we get the following relations for the
ordinary energy density, tangential and radial pressures that will
be helpful to discuss the energy bounds.
\begin{eqnarray}\label{21}
\rho&&=\frac{M e^{-\frac{r^2}{4 \theta }}}{8 \pi ^{3/2} \theta
^{3/2}},\\\label{22} p_{r}&&=-\frac{\text{C}_1+\frac{M(\lambda +8
\pi )erf\left(\frac{r}{2 \sqrt{\theta }}\right)}{4 \pi
}-\frac{M(\lambda +8 \pi ) re^{-\frac{r^2}{4 \theta }}}{4 \pi ^{3/2}
\sqrt{\theta }}}{(\lambda +8 \pi ) r^3},\\\label{23} p_{t}&&=\frac{2
\sqrt{\pi } \left(\frac{4 \pi  \text{C}_1}{\lambda +8 \pi
}+\text{erf}\left(\frac{r}{2 \sqrt{\theta }}\right) M\right
)-\frac{r e^{-\frac{r^2}{4 \theta }} \left(2 \theta
+r^2\right)M}{\theta ^{3/2}}}{16 \pi ^{3/2} r^3}.
\end{eqnarray}

In case of noncommutative geometry with the reference to Lorentzian
distribution, we take the density function as follows
\begin{equation}\label{24}
\rho(r)=\frac{M \sqrt{\theta}}{\pi^{2}(r^{2}+\theta)^{2}},
\end{equation}
where $M$ is a mass which is diffused centralized object such as a
wormhole and $\theta$ is a noncommutative parameter. Comparing
(\ref{16}) and (\ref{24}) and then by solving the resulting
differential equation, we get the following form of shape function:
\begin{equation}\label{25}
b(r)=\frac{(\lambda +8 \pi ) M \left(\left(\theta +r^2\right) \tan
^{-1}\left(\frac{r}{\sqrt{\theta }}\right)-\sqrt{\theta }
r\right)}{2 \pi ^2 \left(\theta +r^2\right)}+C_2,
\end{equation}
where $C_2$ is an integration constant. Again using equation (\ref{25}) in
(\ref{16}) and (\ref{18}), we get a new set of equations which help us to
discuss the energy conditions for the existence of wormhole structure. In
this case, the expressions for energy density, radial and tangential
pressures are given by
\begin{eqnarray}
\rho&&=\frac{\sqrt{\theta } M}{\pi ^2 \left(\theta +r^2\right)^2},\label{26}\\
p_{r}&&=-\frac{C_2+\frac{(\lambda +8 \pi ) M \left(\left(\theta
+r^2\right) \tan ^{-1}\left(\frac{r}{\sqrt{\theta
}}\right)-\sqrt{\theta } r\right)}{2 \pi ^2 \left(\theta +r^2\right)}}{(\lambda +8 \pi ) r^3},\label{27}\\
p_{t}&&=-\frac{\sqrt{\theta } M}{2 \pi ^2 \left(\theta
+r^2\right)^2}+\frac{C_2+\frac{(\lambda +8 \pi ) M
\left(\left(\theta +r^2\right) \tan ^{-1}\left(\frac{r}{\sqrt{\theta
}}\right)-\sqrt{\theta } r\right)}{2 \pi ^2 \left(\theta
+r^2\right)}}{2 (\lambda +8 \pi ) r^3}.\label{28}
\end{eqnarray}

Now we will present the graphical illustration of the obtained shape
functions as well as the conditions that are needed to be fulfilled
for wormhole existence. For this purpose, we take different suitable
choices for the involved free parameters. Firstly, we check the
behavior of shape function $b(r)$ for Gaussian distribution where
the red shift function has been taken as a constant. The left graph
of Figure \textbf{1} indicates the positive increasing behavior of
the shape function and its right graph corresponds to the behavior
of shape function ratio to radial coordinate, i.e., $\frac{b(r)}{r}$
which shows that as the radial coordinate gets larger values, the
ratio $\frac{b(r)}{r}$ approaches to zero, and hence confirms the
asymptotic behavior of shape function. The left part of Figure
\textbf{2} indicates the behavior of $b(r)-r$ which shows that the
wormhole throat for this model is located at $r_0=0.2$ where
$b(r_0)=r_0$. In the right part of this figure, we check the flaring
out condition for this model by plotting $b^{'}(r)$. It shows that
at wormhole throat $r_0=0.2$, clearly the condition $b^{'}(r_0)<1$
is satisfied. The graphical behavior of density function as well as
the null energy conditions $\rho+p_r$ and $\rho+p_t$ are shown in
Figures \textbf{3} and \textbf{4}, respectively. It is clear from
these graphs that the energy density function and the function
$\rho+p_t$ indicate the positive but decreasing behavior versus
radial coordinate while $\rho+p_r$ shows negative and increasing
behavior and hence violates the NEC. Thus it can be concluded that
the obtained wormhole solutions are acceptable in this modified
gravity.
\begin{figure}
\centering \epsfig{file=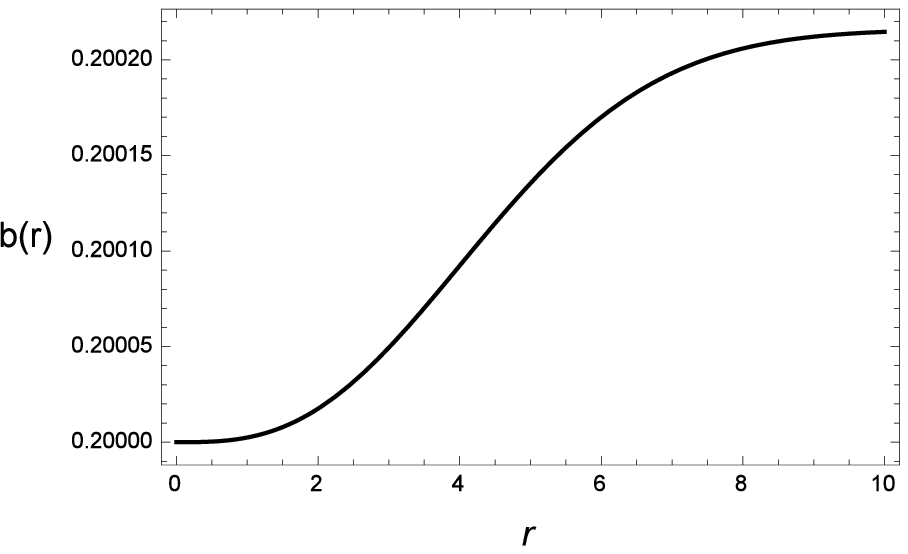, width=.45\linewidth,
height=1.4in}\epsfig{file=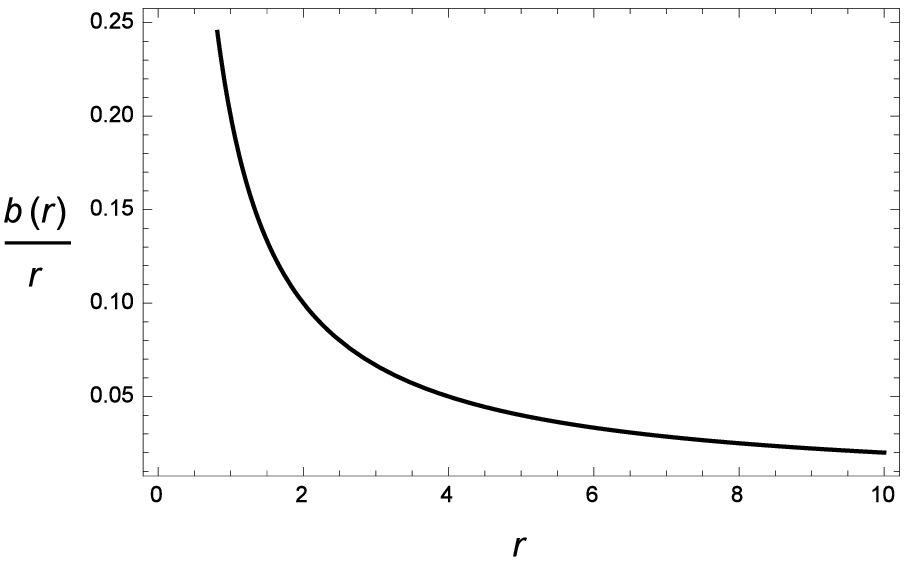, width=.45\linewidth,
height=1.4in} \caption{\label{fig1} This shows the behavior of shape
function $b(r)$ and $\frac{b(r)}{r}$ versus $r$ for Gaussian
distribution. Here, we fix the free parameters as
$\theta=4,~M=0.0001,~C_1=0.2$ and $\lambda=2$.}
\end{figure}
\begin{figure}
\centering \epsfig{file=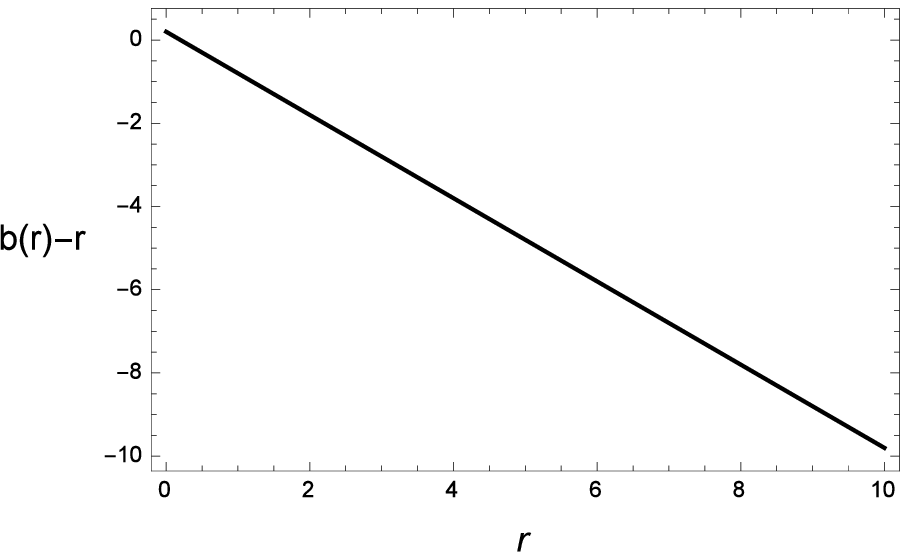, width=.45\linewidth,
height=1.4in}\epsfig{file=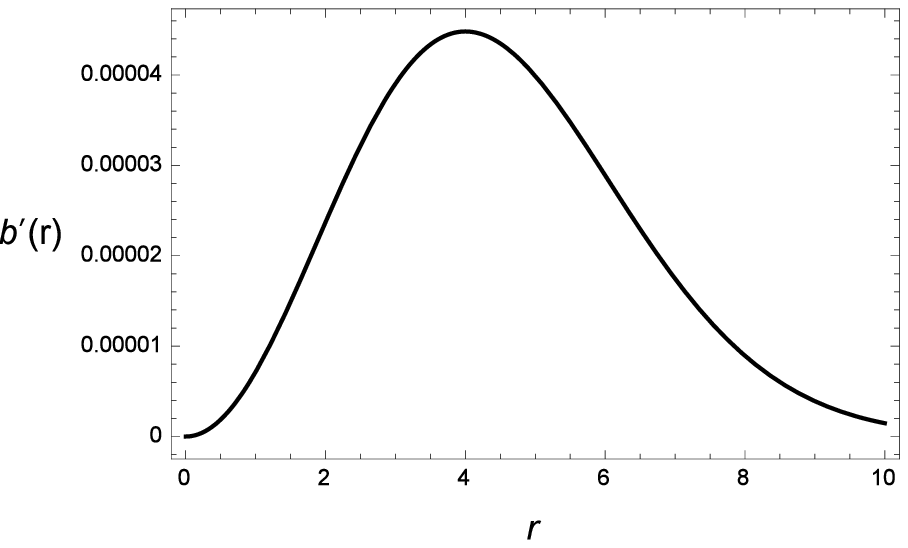, width=.45\linewidth,
height=1.4in} \caption{\label{fig2} This indicates the development
of $b(r)-r$ and $b^{'}(r)$ versus $r$ for Gaussian distribution.
Here, we choose some specific values of the free parameters as
$\theta=4, M=0.0001, C_1=0.2$ and $\lambda=2$.}
\end{figure}
\begin{figure}
\centering \epsfig{file=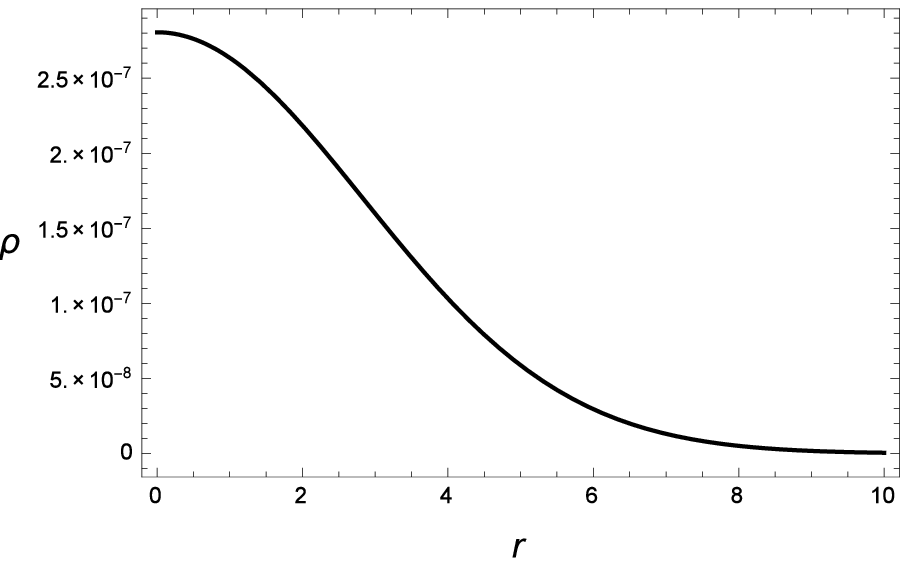, width=.45\linewidth,
height=1.4in}\epsfig{file=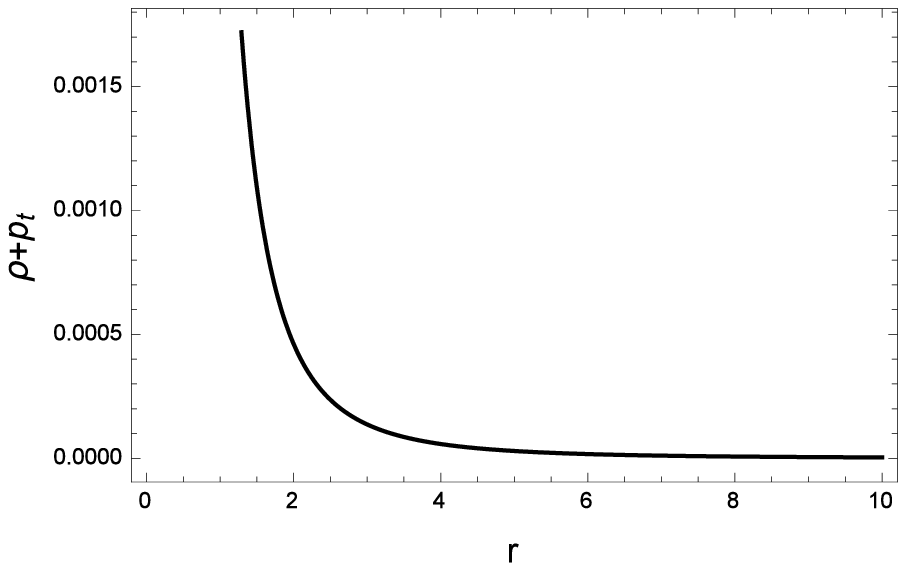, width=.45\linewidth,
height=1.4in} \caption{\label{fig3} This shows the graphical
illustration of $\rho$ and $\rho+p_{t}$ versus $r$ for Gaussian
distribution. Here, we used these values of free parameters
$\theta=4, M=0.0001, C_1=0.2$ and $\lambda=2$.}
\end{figure}
\begin{figure}
\centering \epsfig{file=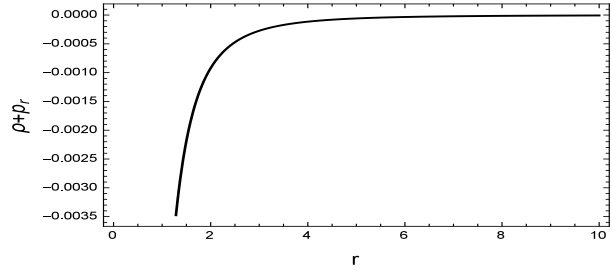, width=.45\linewidth,
height=1.4in}\caption{\label{fig4} This corresponds to the
development of $\rho+p_{r}$ versus $r$ for Gaussian distribution.
Here, we choose $\theta=4, M=0.0001, C_1=0.2$ and $\lambda=2$.}
\end{figure}

In case of wormhole solution with Lorentzian distribution, the graphical
behavior of shape function as well as its corresponding properties are given
in Figures \textbf{5}-\textbf{7}. The left curve of Figure \textbf{5}
corresponds to the behavior of shape function while the right graph shows the
behavior of $\frac{b(r)}{r}$. It is clear from the curves that the shape
function is positive and increasing satisfying the asymptotic flatness
condition as $r\rightarrow0$. Figure \textbf{6} indicates the location of
wormhole throat and the flaring out condition. It is seen that the wormhole
throat is located at $r_0=0.1$ where the function $b(r)-r$ crosses the radial
coordinate axis. Also, at this wormhole throat, the flaring out condition
$b^{'}(r_0)<1$ is satisfied for this case as provided in the right part of
Figure \textbf{6}. The behavior of energy density profile and the functions
$\rho+p_r$ and $\rho+p_t$ is presented in Figures \textbf{7} and \textbf{8}.
These show that the energy density remains positive and increasing with
increasing values of $r$. Similarly, the function $\rho+p_t$ indicates the
positive but decreasing behavior whereas the function $\rho+p_r$ shows the
negative increasing behavior versus $r$. This confirms the violation of NEC
in this case and hence allows the wormhole existence. Thus in both cases, all
necessary and important characteristics of shape function for the wormhole
existence are satisfied and thus it can be concluded that the obtained
solutions are physically viable.
\begin{figure}
\centering \epsfig{file=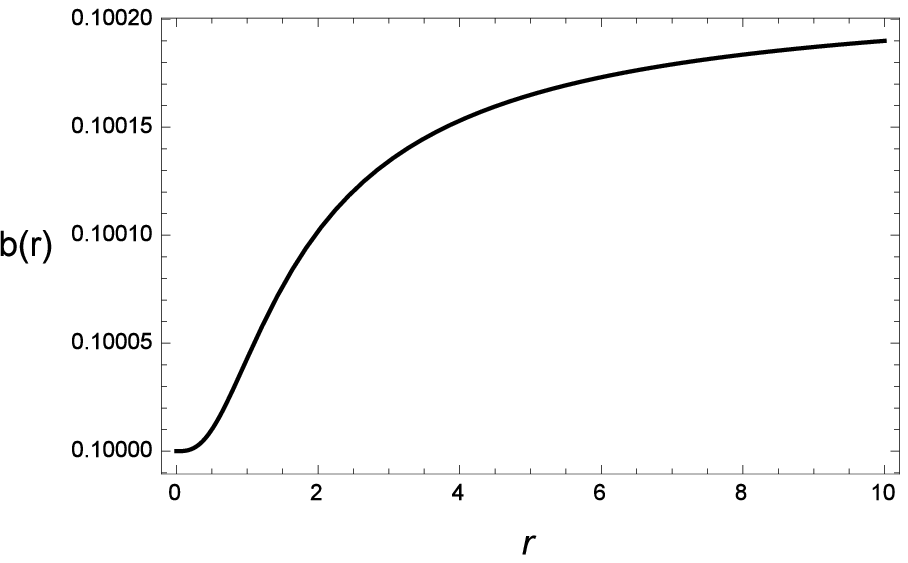, width=.45\linewidth,
height=1.4in}\epsfig{file=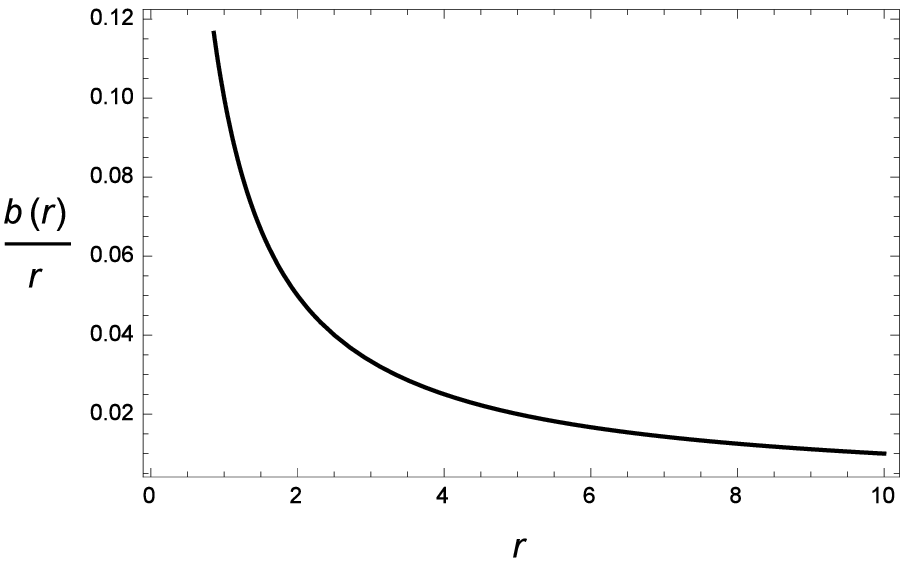, width=.45\linewidth,
height=1.4in} \caption{\label{fig4} This indicates the behavior of
$b(r)$ and $\frac{b(r)}{r}$ versus $r$ for Lorentzian distribution.
Here, we choose the free parameters as $\theta=0.9, M=0.0001,
C_2=0.1$ and $\lambda=2$.}
\end{figure}
\begin{figure}
\centering \epsfig{file=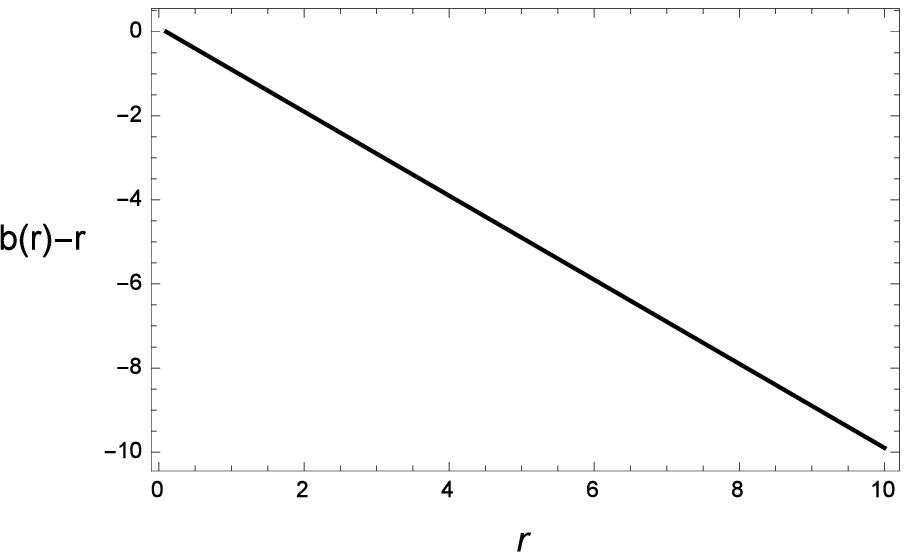, width=.45\linewidth,
height=1.4in}\epsfig{file=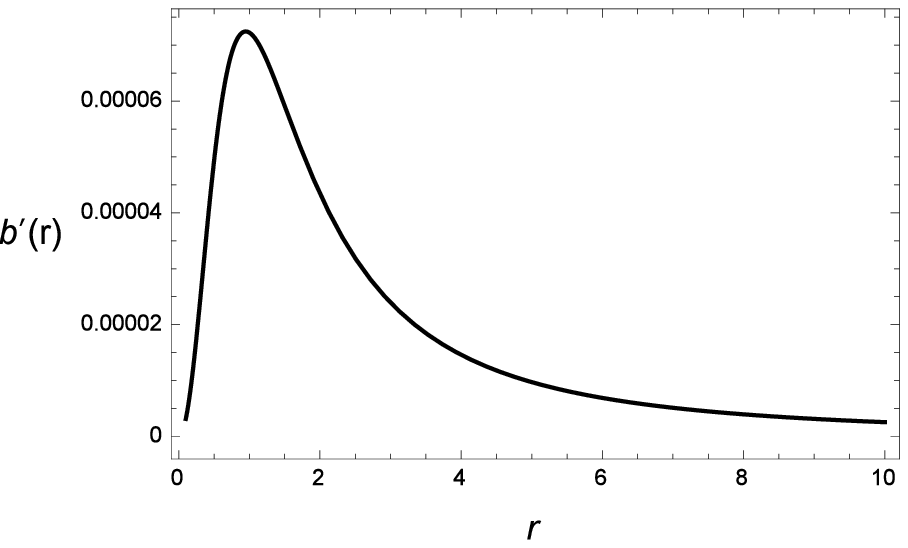, width=.45\linewidth,
height=1.4in} \caption{\label{fig5} This illustrate the development
of $b(r)-r$ and $b^{'}(r)$ versus $r$ for Lorentzian distribution.
Here, we fix the free parameters as $\theta=0.9, M=0.0001, C_2=0.1$
and $\lambda=2$.}
\end{figure}
\begin{figure}
\centering \epsfig{file=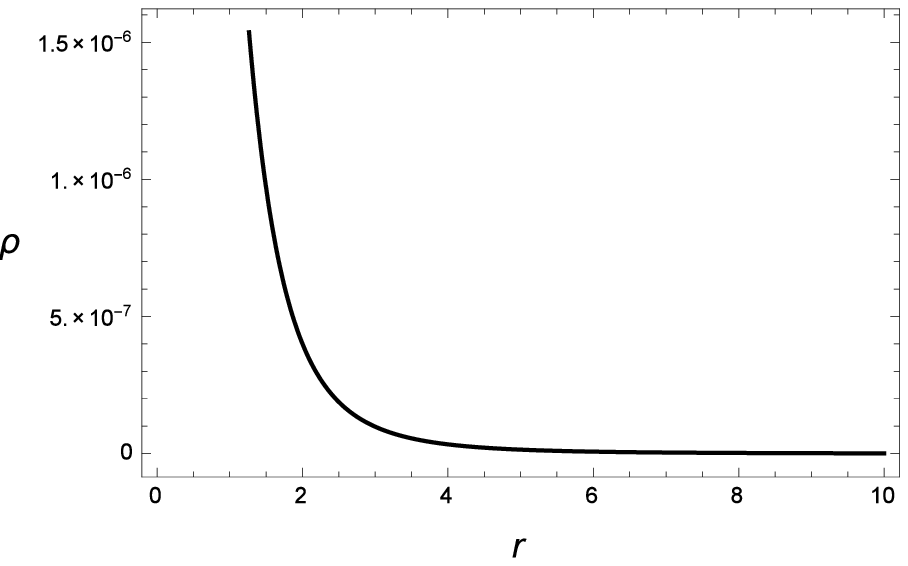, width=.45\linewidth,
height=1.4in}\epsfig{file=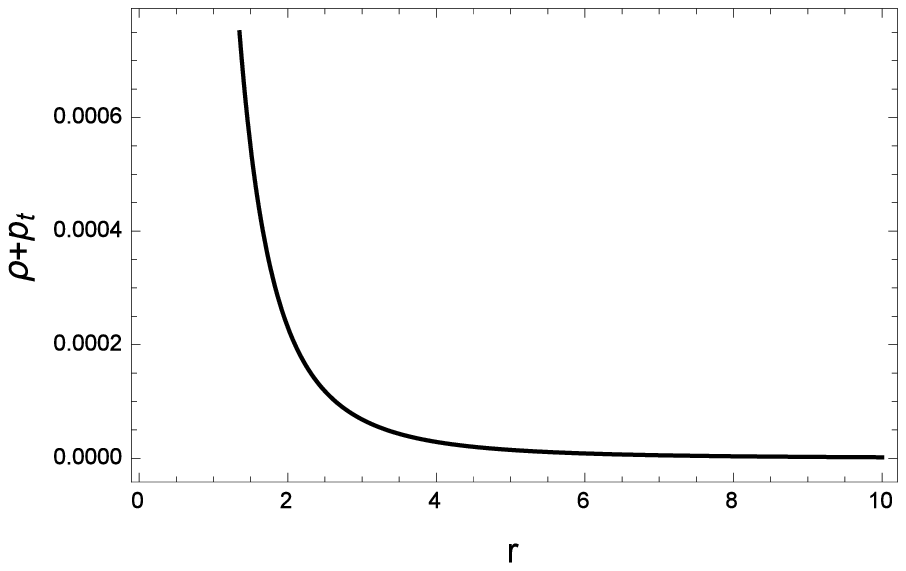, width=.45\linewidth,
height=1.4in} \caption{\label{fig6} This shows graph of $\rho$ and
$\rho+p_{t}$ versus $r$ for Lorentzian distribution. Here,
$\theta=0.9, M=0.0001, C_2=0.1$ and $\lambda=2$.}
\end{figure}
\begin{figure}
\centering \epsfig{file=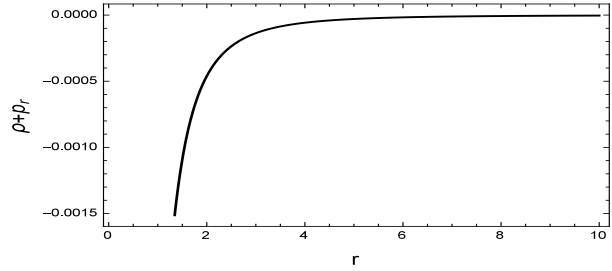, width=.45\linewidth,
height=1.4in}\caption{\label{fig6} This shows $\rho+p_{r}$ versus
$r$ for Lorentzian distribution. Here, we fix $\theta=0.9, M=0.0001,
C_2=0.1$ and $\lambda=2$.}
\end{figure}

\section{Wormhole Solutions: Gaussian and Lorentzian Distributions for $f_{1}(R)=R+\alpha R^{2}+\gamma R^{n}$ Model}

In this segment, we will consider another specific $f_{1}(R)$ model
\cite{23, 24} which is given by the relation
\begin{eqnarray}\label{29}
f_{1}(R)=R+\alpha R^{2}+\gamma R^{n},
\end{eqnarray}
where $\alpha$ and $\gamma$ are arbitrary constants while $n\geq3$. Using the
model (\ref{29}) in Eqs.(\ref{12})-(\ref{14}), we get the following set of
equations for energy density, radial and tangential pressures
\begin{eqnarray}\label{30}
\rho&=&\frac{b'(r)\left(\gamma  2^{n-1} n
(\frac{b'(r)}{r^2})^{n-1}+\frac{4 \alpha b'(r)}{r^2}+1\right)}{r^2(8
\pi+\lambda )},\\\nonumber p_r&=&\frac{1}{4 r^5 (8 \pi +\lambda )
(b'(r))^3}\left(-8 r \alpha b'(r)^3 \left(-8 r b''(r)+b'(r)
\left(12+2 b'(r)\right.\right.\right.\\\nonumber
&-&\left.\left.\left. r b''(r)\right)+2 r^2 b^{(3)}(r)\right)-2^n
(-1+n) n r^5 \gamma \left(\frac{b'(r)}{r^2}\right)^n \left(2
(b'(r))^3+2 \right.\right.\\\nonumber &\times&\left.\left.(-2+n) r^2
(b''(r)^2+(b'(r))^2 \left(-4+8 n-r b''(r)\right)+2 r
b'(r)\right.\right.\\\nonumber &\times&\left.\left. \left(-4 (-1+n)
b''(r)+r b^{(3)}(r)\right)\right)+b(r) \left(2^n n r^4 \gamma
\left(\frac{b'(r)}{r^2}\right)^n \left(2
n\right.\right.\right.\\\nonumber &\times&\left.\left.\left. (-5+4
n) b'(r)^2+2 (-2+n) (-1+n) r^2 b''(r)^2-(-1+n) r b'(r)
\right.\right.\right.\\\nonumber
&\times&\left.\left.\left.\left((-7+8 n) b''(r)-2 r
b^{(3)}(r)\right)\right)+4 (b'(r))^3 \left(24 \alpha  b'(r)+r
\left(-r\right.\right.\right.\right.\\\label{31}
&-&\left.\left.\left.\left.18 \alpha  b''(r)+4 r \alpha
b^{(3)}(r)\right)\right)\right)\right),\\\nonumber p_t&=&-\frac{1}{4
r^5 (8 \pi +\lambda ) (b'(r))^2}\left(2^n n r^4 \gamma
\left(\frac{b'(r)}{r^2}\right)^n \left(b'(r) \left((-5+4 n) b(r)+r
\right.\right.\right.\\\nonumber &\times&\left.\left.\left.\left(4-4
n+b'(r)\right)\right)+2 (-1+n) r (r-b(r)) b''(r)\right)+2 (b'(r))^2
\left(r \left(b'(r)
\right.\right.\right.\\\nonumber&-&\left.\left.\left.\left(r^2-16
\alpha
+4 \alpha  b'(r)\right)+8 r \alpha  b''(r)\right)-b(r) \left(-12 \alpha  b'(r)+r\right.\right.\right.\\
\label{32}&\times&\left.\left.\left. \left(r+8 \alpha
b''(r)\right)\right)\right) \right).
\end{eqnarray}
The comparison of Eqs.(\ref{19}) and (\ref{30}) (Gaussian
distribution) yields the following non-linear differential equation:
\begin{equation}\nonumber
\frac{b'(r) \left(\gamma  2^{n-1} n
\left(\frac{b'(r)}{r^2}\right)^{n-1}+\frac{4 \alpha
b'(r)}{r^2}+1\right)}{(\lambda +8 \pi ) r^2}=\frac{M
e^{-\frac{r^2}{4 \theta }}}{(4 \pi  \theta )^{3/2}},
\end{equation}
which is complicated and hence we solve it numerically for the shape function
$b(r)$.

In the similar way, by comparing Eqs.(\ref{24}) and (\ref{30})
(Lorentzian distribution), we get the following non-linear
differential equation:
\begin{equation}\nonumber
\frac{b'(r) \left(\gamma  2^{n-1} n
\left(\frac{b'(r)}{r^2}\right)^{n-1}+\frac{4 \alpha
b'(r)}{r^2}+1\right)}{(\lambda +8 \pi ) r^2}=\frac{M\sqrt{\theta }
}{\pi ^2 \left(\theta +r^2\right)^2}
\end{equation}
whose analytic solution is also not possible, thus we evaluate the
possible form of shape function by solving this equation
numerically.
\begin{figure}
\centering \epsfig{file=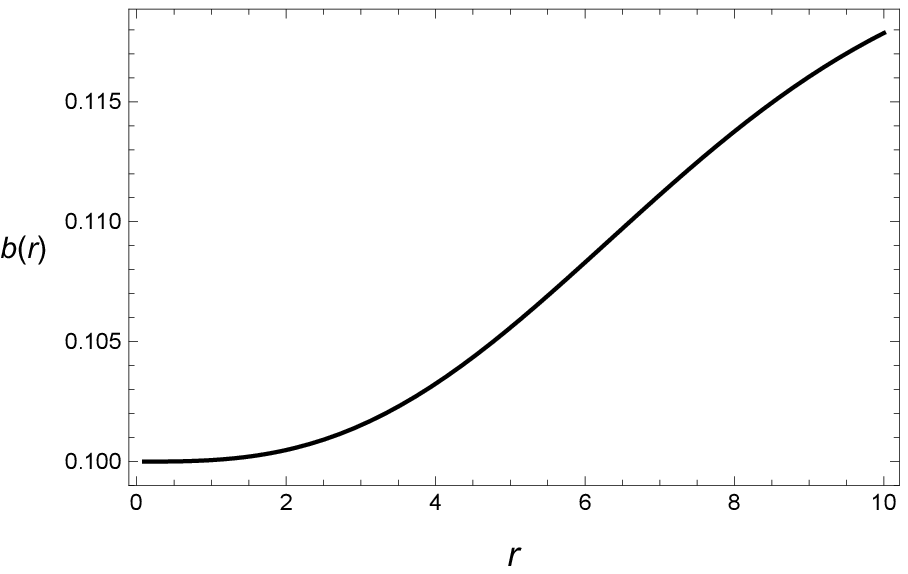, width=.45\linewidth,
height=1.4in}\epsfig{file=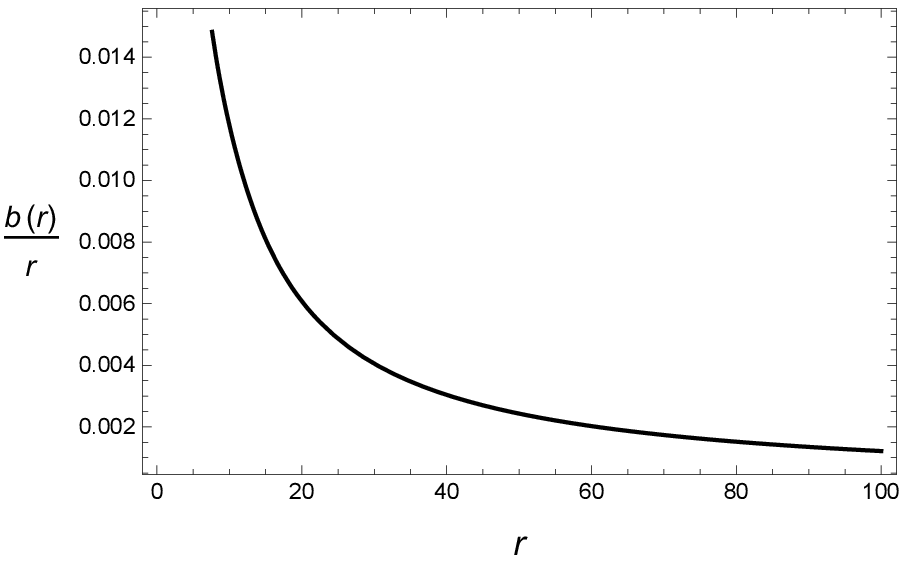, width=.45\linewidth,
height=1.4in} \caption{\label{fig7} This corresponds to the
development of $b(r)$ and $\frac{b(r)}{r}$ versus $r$ for Gaussian
distribution. Here, we take $\theta=10, M=0.01, n=3, \alpha=2.5,
\gamma=2.5$ and $\lambda=2$.}
\end{figure}
\begin{figure}
\centering \epsfig{file=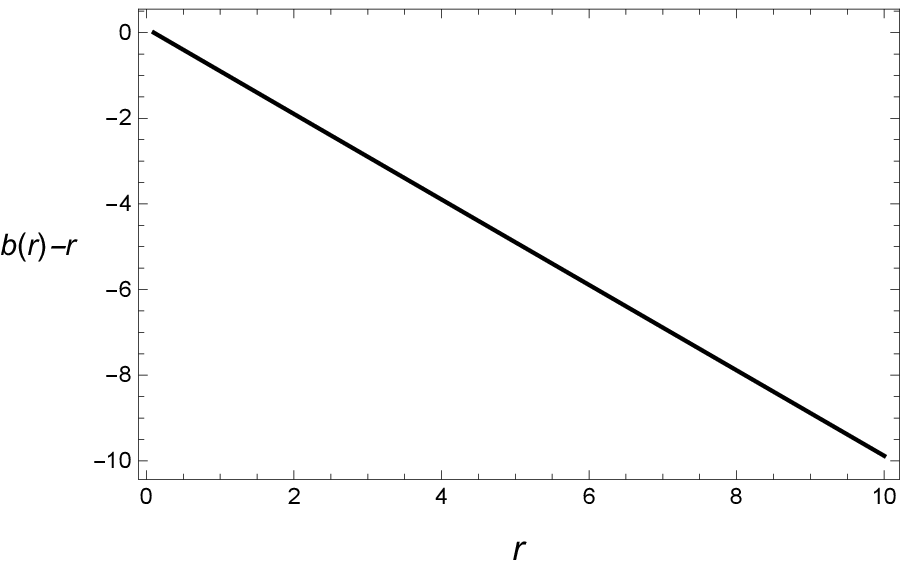, width=.45\linewidth,
height=1.4in}\epsfig{file=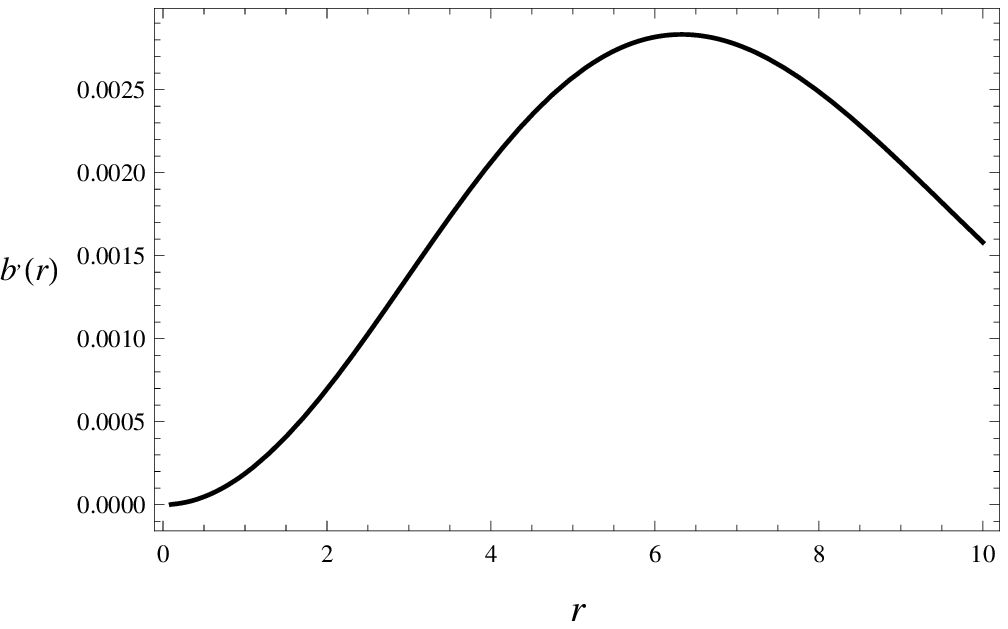, width=.45\linewidth,
height=1.4in} \caption{\label{fig8} This shows the graphs of
$b(r)-r$ and $b^{'}(r)$ versus $r$ for Gaussian distribution. Here,
$\theta=10, M=0.01, n=3, \alpha=2.5, \gamma=2.5$ and $\lambda=2$.}
\end{figure}
\begin{figure}
\centering \epsfig{file=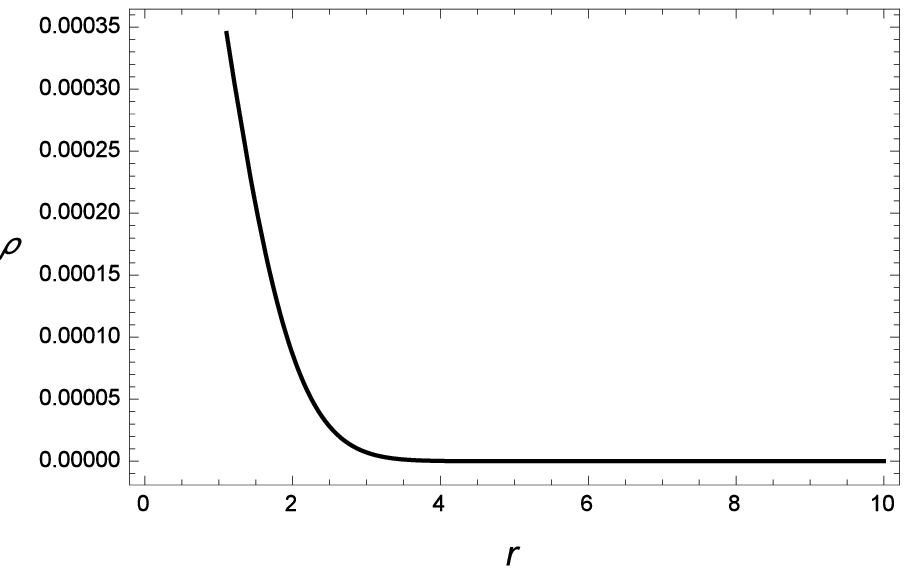, width=.45\linewidth,
height=1.4in}\epsfig{file=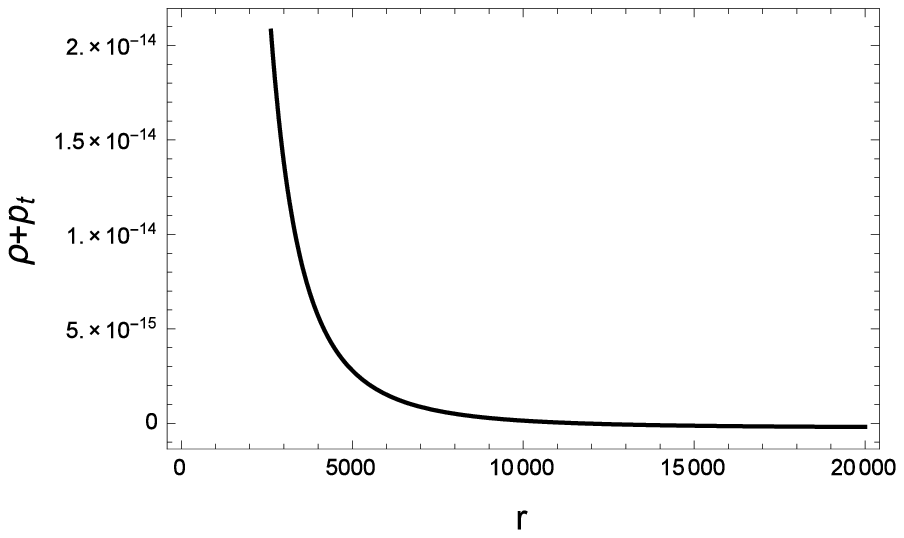, width=.45\linewidth,
height=1.4in} \caption{\label{fig9} This indicates the behavior of
$\rho$ and $\rho+p_{t}$ versus $r$ for Gaussian distribution. Here,
$\theta=0.5, M=0.01, n=3, \alpha=2.5, \gamma=2.5$ and $\lambda=2$.}
\end{figure}
\begin{figure}
\centering \epsfig{file=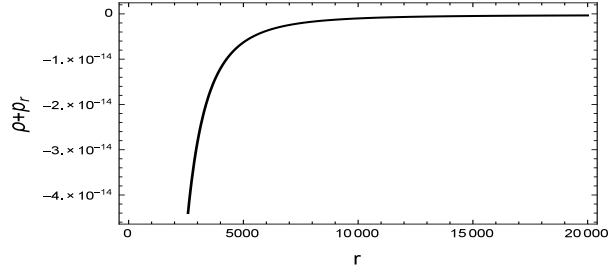, width=.45\linewidth,
height=1.4in}\caption{\label{fig9} This provides the development of
$\rho+p_{r}$ versus $r$ for Gaussian distribution with same values
of the involved parameters.}
\end{figure}

Now we will discuss the behavior of shape functions that are obtained by
numerical approach as well as their corresponding important and necessary
properties for the existence of wormhole structure for both Gaussian and
Lorentizian distributions. For this purpose, we utilize a fixed value $n=3$
for the modified model (\ref{29}) which results in the cubic form given by
$f(R)=R+\alpha R^{2}+\gamma R^{3}$. For the other higher values, i.e., $n>3$,
it is observed that the resulting form of shape function is not physically
viable. For graphical illustration of shape functions and their other
properties, we will take different feasible values of the free parameters.
The left part of Figure \textbf{9} indicates that the shape function remains
positive and increasing for the Gaussian distribution (obtained numerically),
while its right part shows behavior of the function $\frac{b(r)}{r}$ versus
radial coordinate. Clearly, it indicates that as the radial coordinate
increases, the function tends to zero and hence leads to the asymptotic
behavior of shape function. In Figure \textbf{10}, the left curve corresponds
to the function $b(r)-r$ which provides the location of wormhole throat at
$r_0=0.001$ where it cuts the $r$-axis. Its right curve provides information
about the flaring out condition, i.e., $b^{'}(r_0)<1$ which is clearly
compatible at the obtained wormhole throat. Furthermore, the graphical
illustration of energy density, tangential and radial pressures is given in
Figures \textbf{11} and \textbf{12}. The left part of Figure \textbf{11}
corresponds to energy density which shows positive but decreasing behavior
while the right curve shows the graph of $\rho+p_t$ which is also positive
decreasing. Figure \textbf{12} indicates the behavior of $\rho+p_r$ which is
clearly negative and increasing versus $r$ and hence violates the NEC. Thus
all the conditions are satisfied allowing the existence of physically viable
wormhole solution.

\begin{figure}
\centering \epsfig{file=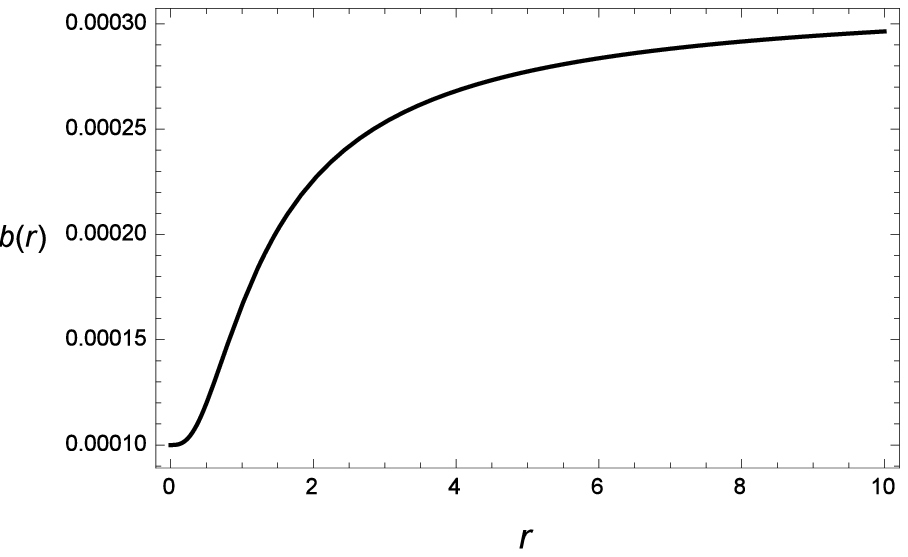, width=.45\linewidth,
height=1.4in}\epsfig{file=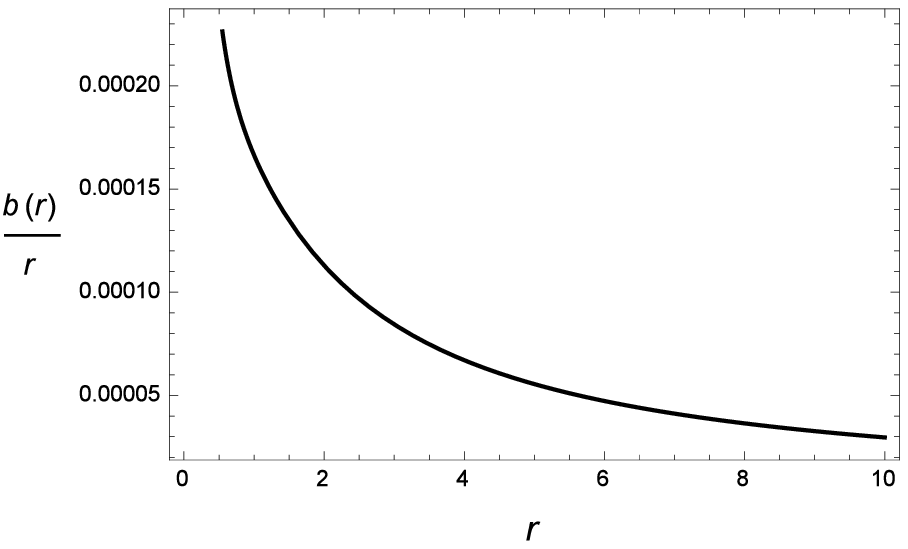, width=.45\linewidth,
height=1.4in} \caption{\label{fig10} This shows the behavior of
$b(r)$ and $\frac{b(r)}{r}$ versus $r$. Here, we choose the free
parameters as $\theta=0.5, M=0.0001, n=3, \alpha=2, \gamma=2$ and
$\lambda=2$.}
\end{figure}
\begin{figure}
\centering \epsfig{file=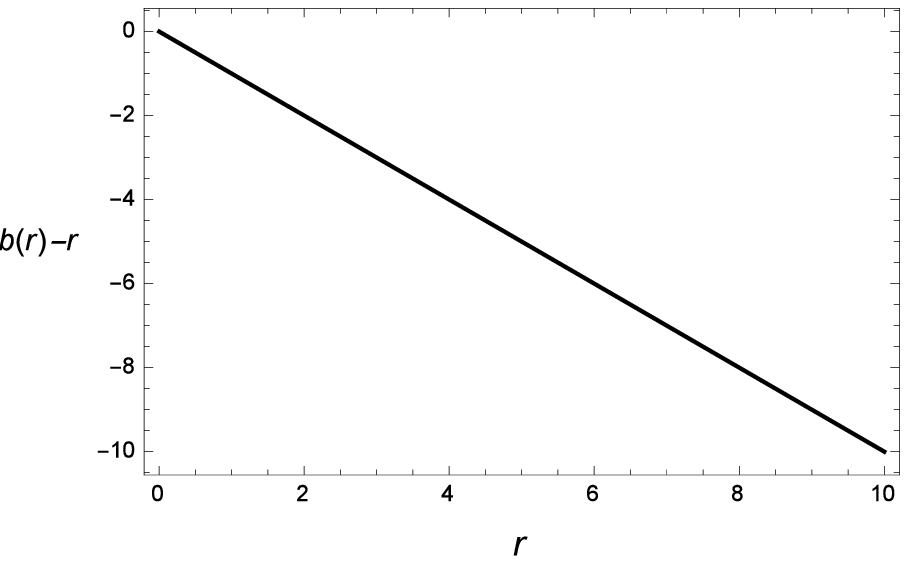, width=.45\linewidth,
height=1.4in}\epsfig{file=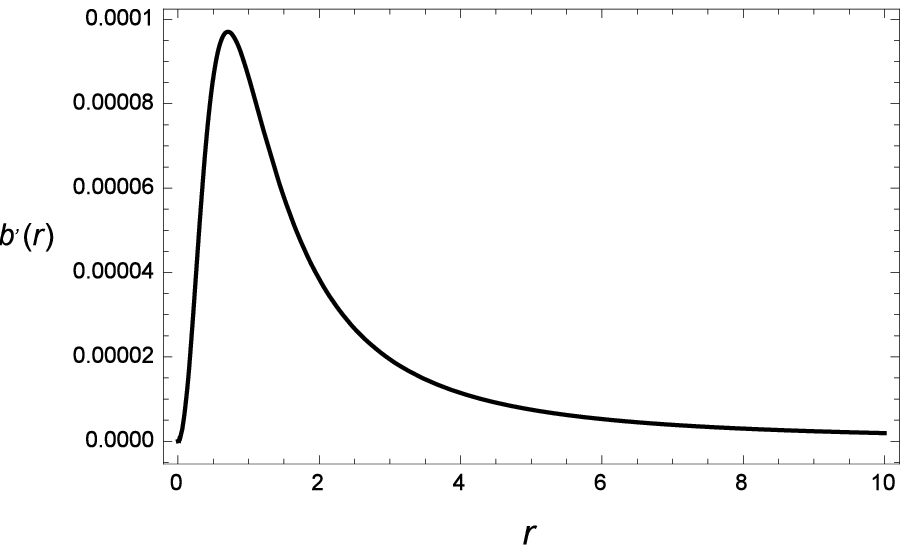, width=.45\linewidth,
height=1.4in} \caption{\label{fig11} This corresponds to the
development of $b(r)-r$ and $b^{'}(r)$ versus $r$ for Lorentzian
distribution with the same choices of free parameters.}
\end{figure}
\begin{figure}
\centering \epsfig{file=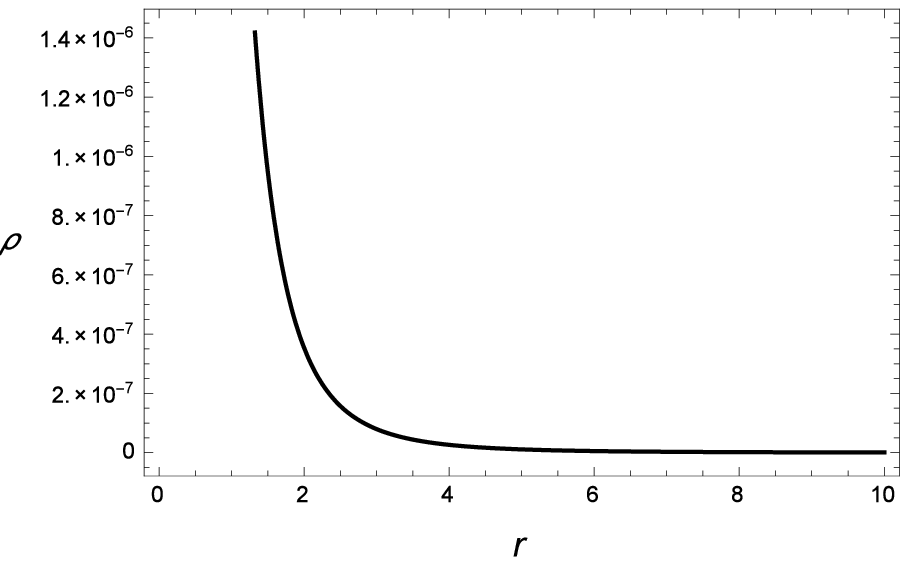, width=.45\linewidth,
height=1.4in}\epsfig{file=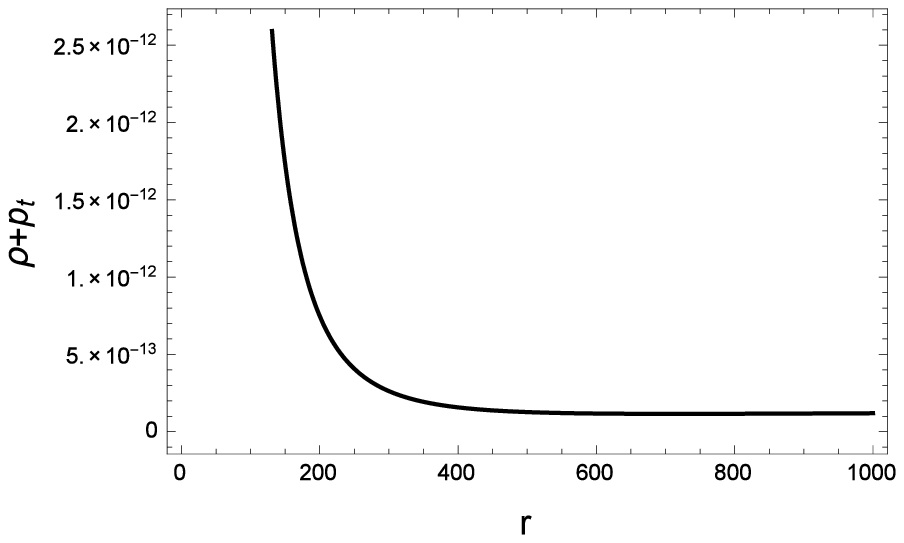, width=.45\linewidth,
height=1.4in} \caption{\label{fig12} This indicates the graphs of
$\rho$ and $\rho+p_{t}$ versus $r$ for Lorentzian distribution with
the same free parameters choices.}
\end{figure}
\begin{figure}
\centering \epsfig{file=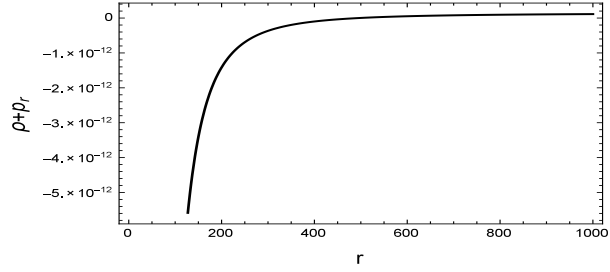, width=.45\linewidth,
height=1.4in}\caption{\label{fig12} This shows the development of
$\rho+p_{r}$ versus $r$ for Lorentzian distribution.}
\end{figure}

Similarly, for the Lorentzian distribution, the graphical behavior of shape
function and its properties like asymptotic behavior, wormhole throat and the
flaring out condition are shown in figures \textbf{13} and \textbf{14}. It
can be easily observed that the obtained shape function is positive
increasing and is compatible with all conditions. Further, the graphs for
resulting density profile, $\rho+p_{t}$ and $\rho+p_{r}$ are given in Figures
\textbf{15} and \textbf{16}, respectively which confirm the violation of NEC
for this wormhole model. Thus it can be concluded that the obtained wormhole
solutions for this cubic polynomial $f_1(R)$ model are physically interesting
for both non-commutative distributions.

\section{Equilibrium Condition}

In this segment, we explore the stability of obtained solutions
using equilibrium conditions in the presence of Gaussian and
Lorentzian distributions of non-commutative geometry. For this
purpose, we take Tolman-Oppenheimer-Volkov equation \cite{17} which
is given by
\begin{equation}\label{33}
\frac{dp_{r}}{dr}+\frac{\sigma^{'}}{2}(\rho+p_{r})+\frac{2}{r}(p_{r}-p_{t})=0,
\end{equation}
where $\sigma(r)=2\Phi(r)$. This equation determines the equilibrium
state of configuration by taking the gravitational, hydrostatic as
well as the anisotropic forces (arising due to anisotropy of matter)
into account. These forces are defined by the following relations:
\begin{equation*}
F_{gf}=-\frac{\sigma^{'}(\rho+p_{r})}{2},\;\;\;\;\;\;\;\;F_{hf}=-\frac{dp_{r}}{dr}, \;\;\;\;\;\;\;\;F_{af}=2\frac{(p_{t}-p_{r})}{r},
\end{equation*}
and thus Eq.(\ref{33}) takes the form given by
\begin{equation*}
F_{af}+F_{gf}+F_{hf}=0.
\end{equation*}
Since we assumed the red shift function as a constant so that $\Phi'(r)=0$,
therefore it leads to $F_{gf}=0$ and hence the equilibrium condition reduces
to the following form:
\begin{equation*}
F_{af}+F_{hf}=0.
\end{equation*}
We shall discuss the stability condition for both exact and
numerical solutions in the presence of both distributions of
non-commutative geometry. Firstly, we calculate $F_{af}$ and
$F_{hf}$ for Gaussian distribution as follows
\begin{eqnarray*}
F_{af}&=&\frac{6 \sqrt{\pi } \left(\frac{4 \pi  \text{C1}}{8 \pi+\lambda  }
+\text{erf}\left(\frac{r}{2 \sqrt{\theta }}\right)M\right)
-\frac{M r e^{-\frac{r^2}{4 \theta }} \left(6 \theta +r^2\right)}{\theta ^{3/2}}}{8 \pi ^{3/2} r^4},\\
F_{hf}&=&\frac{6 \sqrt{\pi } \left(-\frac{4 \pi  \text{C1}}{8 \pi+\lambda }-
 \text{erf}\left(\frac{r}{2 \sqrt{\theta }}\right)M\right)
 +\frac{M r e^{-\frac{r^2}{4 \theta }} \left(6 \theta +r^2\right)}{\theta ^{3/2}}}{8 \pi ^{3/2}
 r^4},
\end{eqnarray*}
while for Lorentizian distribution, these are given by
\begin{eqnarray*}
F_{af}&=&\frac{6 \pi ^2 C_2 \left(\theta +r^2\right)^2-\sqrt{\theta
} (8 \pi+\lambda) M r \left(3 \theta +5 r^2\right) +3M (8
\pi+\lambda) \left(\theta +r^2\right)^2 \tan
^{-1}\left(\frac{r}{\sqrt{\theta }}\right)}{2 \pi ^2 (8 \pi+\lambda)
 r^4 \left(\theta +r^2\right)^2},\\
F_{hf}&=&\frac{-6 \pi ^2 C_2 \left(\theta +r^2\right)^2+\sqrt{\theta
} (8 \pi+\lambda) M r \left(3 \theta +5 r^2\right) -3M (8
\pi+\lambda) \left(\theta +r^2\right)^2 \tan
^{-1}\left(\frac{r}{\sqrt{\theta }}\right)}{2 \pi ^2 (8 \pi+\lambda)
r^4 \left(\theta +r^2\right)^2}.
\end{eqnarray*}
The graphical behavior of these forces is given in Figures
\textbf{17} and \textbf{18}. The left graph indicates the behavior
of these forces for Gaussian distribution while the right graph
corresponds to Lorentzian distribution for simple $f_1(R)$ model. It
is clear from the graph that both these forces show the same but
opposite behavior and hence cancel each other's effect and thus
leaving a stable wormhole configuration.
\begin{figure}
\centering \epsfig{file=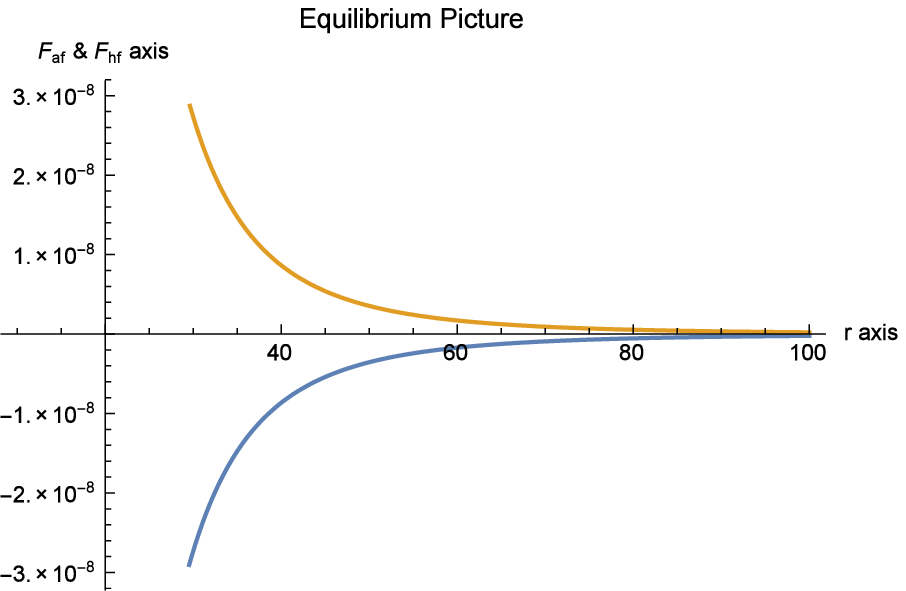, width=.45\linewidth,
height=1.4in}\epsfig{file=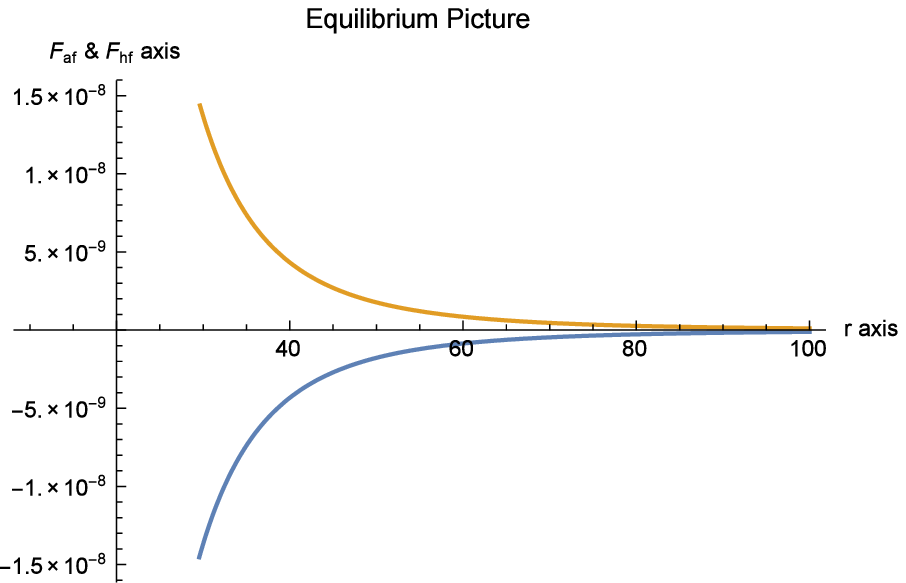, width=.45\linewidth,
height=1.4in} \caption{\label{fig13} This shows the graphical
illustration of $F_{af}$ and $F_{hf}$ forces versus $r$ for Gaussian
and Lorentizian distribution when $f_1(R)=R$ in the left and right
panels, respectively. Here, $\theta=0.5, M=0.01, n=3, \alpha=2.5,
\gamma=2.5$ and $\lambda=2$}
\end{figure}
Similarly, we investigate the stability of numerical solutions for
modified cubic $f(R)$ model using both the Gaussian and Lorentzian
distributions. The graphical behavior of resulting forces is given
by Figure \textbf{18}. Its left part corresponds to behavior of
these forces for Gaussian distribution whereas the right graph
provides the behavior for Lorentzian distribution which clearly
indicates that these forces are also balancing each other's effect
and thus leading to a stable wormhole structure.
\begin{figure} \centering
\epsfig{file=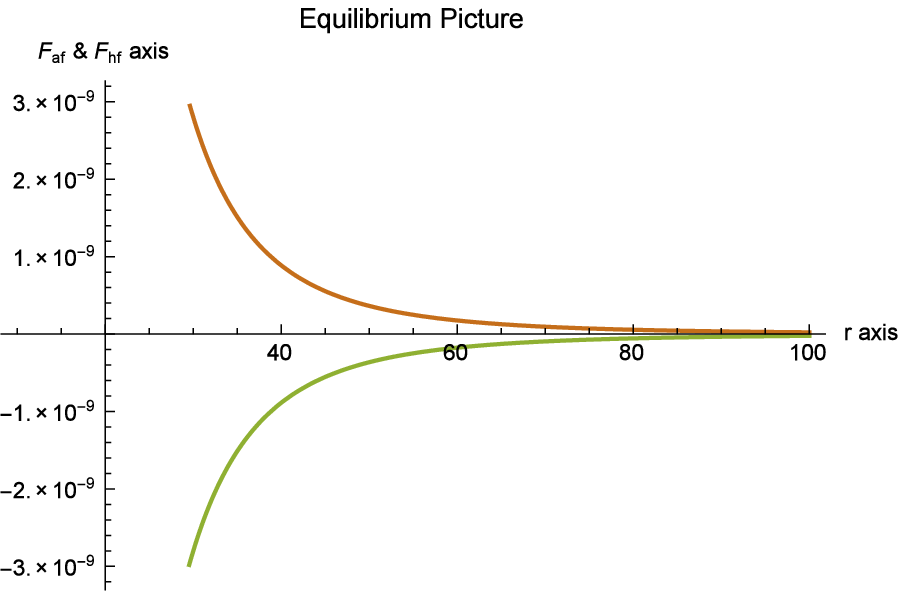, width=.45\linewidth,
height=1.4in}\epsfig{file=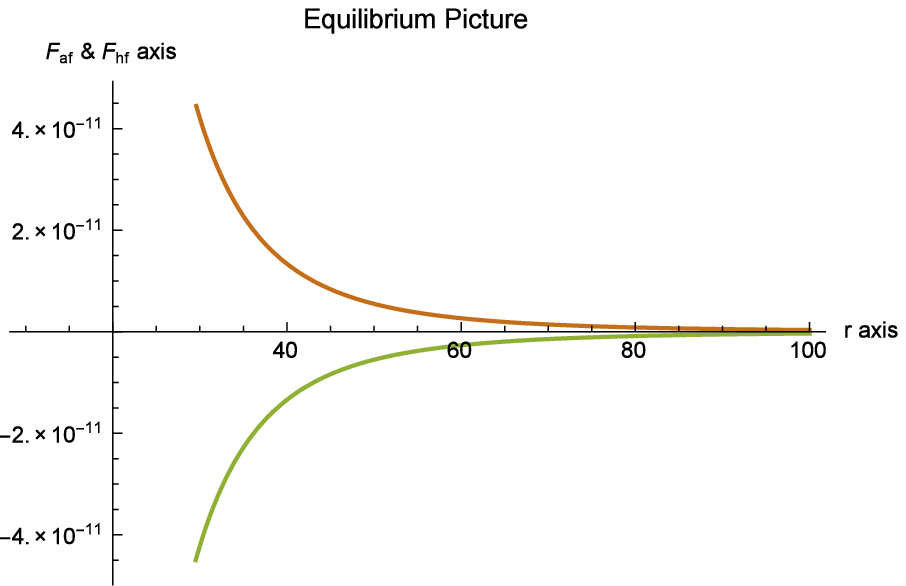, width=.45\linewidth,
height=1.4in}\caption{\label{fig18} This shows the development of
$F_{af}$ and $F_{hf}$ forces versus $r$ for Gaussian and Lorentzian
distributions when $f(R)=R+\alpha R^{2}+\gamma R^{3}$. Here, we fix:
$\theta=0.5, M=0.01, n=3, \alpha=2.5, \gamma=2.5$ and $\lambda=2$.}
\end{figure}

\section{Gravitational Lensing effect of Wormhole for Simple $f(R)=R$ model}

In this section, we will explore the possible detection of
traversable wormhole through gravitational lensing phenomena. For
this purpose, we consider the static spherical symmetric metric
involving $x=\frac{r}{2M}$ representing the radius in Schwarzschild
units and is given by
\begin{eqnarray}\label{1n}
ds^2=-A(x)dt^2+B(x)dx^2+C(x)(d\theta^2+\sin^2\theta d\phi^2).
\end{eqnarray}
Here the closest path taken by the light ray is $\hat{x}=\frac{\hat{r}}{2M}$.
Here we will consider the obtained exact form of shape function in case of
simple linear $f_1(R)=R$ model (section \textbf{III}). The integration of
this shape function from wormhole throat $r_{0}$ to $r$ is given as follows
\begin{equation}\label{2n}
b(r)=\frac{(\lambda+8\pi)M\sqrt{\theta}}{2\pi^2}[\frac{1}{\sqrt{\theta}}
\arctan(\frac{r}{\sqrt{\theta}})-\frac{r}{(r^2+\theta)}-\frac{1}{\sqrt{\theta}}
\arctan(\frac{r_{0}}{\sqrt{\theta}})-\frac{r_{0}}{(r_{0}^2+\theta)}]+r_{0}.
\end{equation}
Here clearly the coupling constant satisfies $\lambda\neq-8\pi$.
Basically, we consider the form of static spherically symmetric
wormhole metric (\ref{5}) where $e^{2\Phi(r)}=(\frac{r}{b_0})^m$
where $b_0$ is an integration constant while $m=2(v^{\phi})^2$,
where $v^{\phi}$ indicates the rotational velocity. In \cite{25}, it
is pointed out that $m=0.000001$ which is very small values (nearly
zero) and hence leaving the red shift function as a constant (as we
assumed in previous sections). The comparison of these metrics leads
to the following relations:
\begin{eqnarray}
A(x)=(\frac{r}{b_0})^m,\quad B(x)=(1-\frac{b(r)}{r})^{-1}, \quad
C(x)=r^2.
\end{eqnarray}
The deflection angle for light ray is given by
\begin{equation}\label{}
\alpha(\hat{x})=-\ln(\frac{2d}{3}-1)-0.8056+I(\hat{x}).
\end{equation}
Here $d$ represents the mouth of wormhole because of exterior
Schwarzschild line element while the internal metric contribution is
provided by $I(\hat{x})$ which determines that the closest path
taken by the ray of light is bigger than the wormhole mouth. This is
defined by the relation:
\begin{equation}
I(\hat{x})=\int^{\infty}_{\hat{x}}\frac{2\sqrt{B(x)dx}}{\sqrt{C(x)\sqrt{\frac{C(x)A(\hat{x})}{C(\hat{x})A(x)}}-1}}.
\end{equation}
In our case, this integral take the following form:
\begin{equation}
I(\hat{x})=\int^{d}_{\hat{x}}G(x)dx
\end{equation}
representing the closest approach for the light ray to be inside the
wormhole mouth. Here the function $G(x)$ is given by
\begin{equation}
G(x)=\frac{2}{\sqrt{x^2[1-\frac{1}{x}(\frac{\lambda+8\pi}{4\pi^2})(\arctan(\frac{x}{\sqrt{\theta}})
-\frac{x\sqrt{\theta}}{(x^2+\theta)}-\arctan(\frac{x_{0}}{\sqrt{\theta}})+\frac{x_{0}\sqrt{\theta}}{(x^2_{0}+\theta)})
+x_{0}]}\sqrt{\frac{x^{2-m}}{\hat{x}^{2-m}}-1}}.
\end{equation}
In order to investigate the convergence/divergence of this integral,
we can redefine the variable as $y=\frac{x}{\hat{x}}$ for the sake
of simplicity in calculations. Thus the integral takes the following
form
\begin{equation}
I(\hat{x})=\int^{d/\hat{x}}_{1}\frac{2}{\sqrt{[1-\frac{1}{\hat{x}y}(\frac{\lambda+8\pi}{4\pi^2})(\arctan(\frac{\hat{x}y}{\sqrt{\theta}})
-\frac{\hat{x}y\sqrt{\theta}}{(\hat{x}^2y^2+\theta)}-\arctan(\frac{\hat{x}y_{0}}{\sqrt{\theta}})+\frac{\hat{x}y_{0}\sqrt{\theta}}{(\hat{x}^2y_{0}^2+\theta)})
+\hat{x}y_{0}]}\sqrt{y^{4-m}-y^2}}.
\end{equation}
In the integrand of the above integral, we can assume that
$H(y)=f(y)(y^{4-m}-y^2)$, where
\begin{equation}
f(y)=1-\frac{1}{\hat{x}y}(\frac{\lambda+8\pi}{4\pi^2})[\arctan(\frac{\hat{x}y}{\sqrt{\theta}})
-\frac{\hat{x}y\sqrt{\theta}}{(\hat{x}^2y^2+\theta)}-\arctan(\frac{\hat{x}y_{0}}{\sqrt{\theta}})+\frac{\hat{x}y_{0}\sqrt{\theta}}{(\hat{x}^2y_{0}^2+\theta)}]
+\hat{x}y_{0}.
\end{equation}
Taylor's series can be used to expand the function $H(y)$ around
$y=1$ as follows
\begin{equation}
H(y)=(2-m)f(1)(y-1)+[\frac{1}{2}(5-m)(2-m)f(1)+(2-m)f'(1)](y-1)^2+O(y-1)^3.
\end{equation}
Here we truncate the Taylor's expansion up to second-order where
$O(y-1)^3$ indicates the cubic and higher-order terms of factor
$(y-1)$. It can be easily observed that the integral $I(\hat{x})$
converges or diverges because of the leading term in the above
expression. Integral can be convergent if the first $(y-1)^{1/2}$
leads the expression where $g(1)\neq0$. If $g(1)=0$, then second
term will lead the expression and whose integration will be
$\ln(y-1)$. Since $y=1$, therefore it turns out be undefined there
and hence the integral diverges. If we choose the nearest approach
of light ray as the wormhole throat, i.e., $\hat{r}=r_{0}$, then
consequently, we have $y_{0}=\frac{x_{0}}{\hat{x}}$ and thus
$y_{0}=1$. Using these values in $f(y)$, it can be easily verified
that $f(1)=0$. Hence a photon sphere with radius $\hat{r}$ (closest
path taken by light ray) equal to throat radius $r_0$, can be found.

\section{Conclusions}

The existence and construction of wormhole solutions in GR with some exotic
matter has always been of great interest for the researchers. The presence of
exotic matter is one of the most important requirement for wormhole
construction as it leads to NEC violation and hence permits the wormhole
existence. In case of modified theories, construction of wormholes has become
more fascinating topic as these include the effective energy-momentum tensor
that violates NEC without inclusion of any exotic matter separately. In the
present paper, we have constructed spherically symmetric wormhole solutions
in the presence of two interesting Gaussian and Lorentzian distributions of
non-commutative geometry in $f(R,T)$ modified gravity. For this purpose, in
order to make system of equations closed, we assumed the function
$f(R,T)=f_1(R)+\lambda T$ with two different forms of $f_1(R)$, i.e., the
linear form $f_1(R)=R$ and $f_{1}(R)=R+\alpha R^{2}+\gamma R^{n},~n\geq3$.

Firstly, we talked about the possible wormhole construction for the linear
$f_1(R)$ model with both Gaussian and Lorentzian distributions. For
Lorentzian and Gaussian distribution, we found the exact solution. In order
to examine the physical behavior of these obtained solutions, we plotted
$b(r)$ versus radial coordinate. It is observed that shape functions show
positive increasing behaviors for both these non-commutative distributions.
Further we found the location of wormhole throats and analyzed some important
characteristics of the shape functions namely asymptotic behavior, the
flaring out condition and the violation of NEC using graphs. This discussion
has been given in Figures \textbf{1}-\textbf{8}. It is concluded from these
graphs that the obtained shape functions show asymptotic behavior, i.e.,
$\frac{b(r)}{r}\rightarrow0$ as $r\rightarrow\infty$. Also, for both cases,
wormhole throats are located at $r_0=0.2$ and $r_0=0.1$. Furthermore, the
obtained shape functions are compatible with the flaring out condition and
NEC as the function $\rho+p_r$ indicated negative behavior for both
distributions. Thus the obtained solutions are viable permitting wormhole
to exist in non-commutative $f(R,T)$ gravity.

Secondly, we checked the wormhole existence for model
$f_{1}(R)=R+\alpha R^{2}+\gamma R^{n},~n\geq3$ by taking both
non-commutative distributions into account. In this case, we
obtained very complicated non-linear differential equations for
$b(r)$ whose analytic solutions are not possible, therefore we
solved them numerically. It is worthwhile to mention here that we
fixed $n=3$ for numerical solutions and their graphical behaviors as
it is found that for $n>3$, the obtained solutions are not
physically interesting (not meeting the necessary criteria for
wormhole existence). In the left parts of Figure \textbf{9} and
\textbf{13}, it is shown that the numerical solutions for $b(r)$
indicate increasing positive behavior. Other necessary conditions
like asymptotic behavior of shape function, flaring out condition as
well as NEC have been given in Figures \textbf{9}-\textbf{16}. The
wormhole throat for solutions in both distributions are located at
$r_0=0.001$. Also, $\rho+p_r$ shows negative behavior and hence NEC
is incompatible for this solution. Thus it is concluded that all the
conditions are satisfied for the chosen specific values of free
parameters and hence the obtained wormhole solutions are viable. It
is also interesting to mention here that for a different selection
of free parameters $\theta,~M,~\lambda$ etc. (other than the used
values in the present paper), all the functions show a similar
graphical behavior as presented in the Figures. Thus all the
necessary conditions for wormhole existence will also be satisfied
in these cases and hence the wormhole solution still exist.

Further, we examined the stability of obtained solutions using equilibrium
condition given by Tolman-Oppenheimer-Volkov equation. Here we explored the
stability for both models of $f_1(R)$ in the presence of Gaussian and
Lorentzian distributions. After evaluating the possible expressions of
anisotropic and hydrostatic forces for these cases, we examined them
graphically as shown in Figures \textbf{17} and \textbf{18}. It can be easily
observed from the graphs that these forces are almost equal in magnitude but
opposite in behavior, therefore canceling each other's effect and hence
leaving a balanced final wormhole configuration. Furthermore, we explored the
possible detection of photon sphere at wormhole throat. For this purpose, we
followed the procedure given in reference \cite{25} and explored the
convergence of deflection angle. It is observed that for the obtained exact
solution for $f_1(R)=R$, the resulting integral diverges at wormhole throat
and hence it is concluded that a photon sphere with radius $r_0$ (closest
path taken by light ray) equal to throat radius, can be detected.

\end{document}